\documentclass[12pt,english,a4paper]{iopart}
\usepackage[T1]{fontenc}
\usepackage[latin9]{inputenc}
\usepackage{babel}
\usepackage{units}
\usepackage{graphicx}
\usepackage[unicode=true, 
 bookmarks=true,bookmarksnumbered=false,bookmarksopen=false,
 breaklinks=false,pdfborder={0 0 1},backref=false,pagebackref=false,
 colorlinks=false]
 {hyperref}
\hypersetup{pdftitle={Reconnection and merging of positive streamers in air},
 pdfauthor={S. Nijdam}}
 
\usepackage{iopams}
\usepackage{setstack}

\makeatletter

\usepackage{bm}

\makeatother

\begin{document}

\title{Reconnection and merging of positive streamers in air}

\author{S~Nijdam$^{1}$, C~G~C~Geurts$^{1}$, E~M~van~Veldhuizen$^{1}$
and U~Ebert$^{1,2}$}

\address{$^{1}$ Eindhoven University of Technology, Dept.\ Applied Physics\\
 P.O. Box 513, 5600 MB Eindhoven, The Netherlands}

\address{$^{2}$ Centrum Wiskunde \& Informatica, Amsterdam, The Netherlands}

\ead{s.nijdam@tue.nl}
\begin{abstract}
Pictures show that streamer or sprite discharge channels emerging
from the same electrode sometimes seem to reconnect or merge though
their heads carry electric charge of the same polarity; one might
therefore suspect that reconnections are an artifact of the two-dimensional
projection in the pictures. Here we use stereo-photography to investigate
the full three-dimensional structure of such events. We analyze reconnection,
possibly an electrostatic effect in which a late thin streamer reconnects
to an earlier thick streamer channel, and merging, a suggested photoionization
effect in which two simultaneously propagating streamer heads merge
into one new streamer. We use four different anode geometries (one
tip, two tips, two asymmetric protrusions in a plate, and a wire),
placed 40~mm above a flat cathode plate in ambient air. A positive
high voltage pulse is applied to the anode, creating a positive corona
discharge. This discharge is studied with a fast ICCD camera, in many
cases combined with optics to enable stereoscopic imaging. We find
that reconnections as defined above occur frequently. Merging on the
other hand was only observed at a pressure of 25~mbar and a tip separation
of 2~mm, i.e., for a reduced tip distance of $p\cdot d=50\:\mu\mathrm{m\, bar}$.
In this case the full width at half maximum of the streamer channel
is more than 10 times as large as the tip separation. At higher pressures
or with a wire anode, merging was not observed. 
\end{abstract}

\submitto{\JPD}

\maketitle

\section{Introduction}

Streamers penetrate into undervolted gaps due to space charges and
local field enhancement at their heads~\cite{Ebert2006a}; frequently
they break up into trees with many branches. Streamer branches stretching
out from the same electrode carry head charges of equal polarity and
repel each other electrostatically. Such self avoiding behaviour is
incorporated in phenomenological dielectric breakdown models, see,
e.g. \cite{Niemeyer1984,Niemeyer1989,Pasko2001,Akyuz2003}. On the
other hand, streamers and leaders emerging from oppositely charged
electrodes carry opposite head charges; therefore when propagating
towards the opposite electrode, they attract each other electrostatically
and frequently join one another along the way; this is seen, e.g.,
in lamp ignition~\cite{Sobota2008,Czichy2008} as well as in the
counter leaders stretching from tall structures upwards towards an
approaching lightning leader~\cite{Bazelyan2008}.

However, there are three recent observations \cite{Briels2006,Grabowski2005,Cummer2006}
that seem to violate this scheme: streamer or sprite channels emerging
from the same electrode do not repel each other, but they seem to
merge or reconnect.

The first type of events was reported by Briels \textit{et al.}~\cite{Briels2006}
in pulsed power experiments in a needle-to-plane electrode geometry.
Here thick and thin positive streamers emerged from the needle anode.
The thick ones are much faster than the thin ones (as quantified in~\cite{Briels2006,Briels2008})
and start somewhat earlier, they reach the cathode plate and then
seem to attract thin streamers that arrive later. The streamers seem
to approach the earlies ones in an almost perpendicular direction.
Such an event is shown in figure~6 of~\cite{Briels2006} and will
be called reconnection; a similar event is shown in figure~\ref{fig:onetipStereoReconnection}
below and in figure~7 of~\cite{Winands2006} and figure~10 of~\cite{Winands2008a}.
A physical mechanism for such an event is given in \cite{Briels2006}:
the originally positive thick streamer channel charges negatively
after connecting to the cathode and therefore attracts the late positive
streamer electrostatically.

The second type of events was seen by Grabowski \emph{et al.~}\cite{Grabowski2005}
and Winands \textit{et al.}~\cite{Winands2006} in pulsed power experiments
of either positive or negative polarity: many streamers emerged from
a wire electrode, and sometimes two almost parallel streamers seemed
to merge into a single one while propagating away from the wire. Such
an event is shown in figure~5B of~\cite{Grabowski2005} and will
be called merging; a similar event is shown in figure~\ref{fig:Wire-plateNoMerging}
below. It was also reported orally for the experiments discussed in~\cite{Winands2006,Winands2008a}.
A physical mechanism for such merging was recently proposed by Luque
\textit{et al.}~\cite{Luque2008b}: the non-local photoionization
reaction in nitrogen-oxygen mixtures like air could generate so much
ionization in the space between the streamer heads that the heads
merge despite their electrostatic repulsion; the mechanism was demonstrated
in three-dimensional computations of two streamers for varying gas
density and nitrogen-oxygen ratio.

A third type of event was reported by Cummer \textit{et al.}~\cite{Cummer2006}
in high speed images of sprite discharges above thunderclouds; the
figures are reproduced in figure~\ref{fig:Sprite_Cummer} below.
Here sprite channels propagating downwards seem to connect to each
other, often accompanied by a bright spot. Sprite discharges have
been established to be large versions of streamers at low gas densities,
related to each other by similarity laws \cite{Pasko2007,Briels2008b,Stenbaek-Nielsen2008}.
This phenomenon looks quite similar to the reconnection described
above, however, the sprite streamers do not reach any electrodes to
change the channel polarity.

These three events were imaged with normal photography, i.e., in a
two-dimensional projection of the full three-dimensional events. Therefore
it is impossible to determine from the figures whether two streamer
branches really do join, or whether they pass behind each other, and
only the statistical analysis of many pictures can lead to such a
conclusion. However, the true three-dimensional event can be reconstructed
from stereo photography. This method was recently developed for streamers
by Nijdam \textit{et al.}~\cite{Nijdam2008} where it has been shown
that such a method can be very successful in reconstructing the complete
3D streamer tree structure for streamer discharges with a limited
number of streamer channels (less than 50). In the paper~\cite{Nijdam2008},
one example of an event is given that looks like a streamer loop or
reconnection in normal photography, but where the channels turn out
to pass behind each other in the stereo photography. In the present
paper, stereo photography is applied to several situations where streamers
appear to reconnect or merge.

\newpage{}

\section{Experimental set-up}

\begin{figure}
\includegraphics[clip,width=9cm]{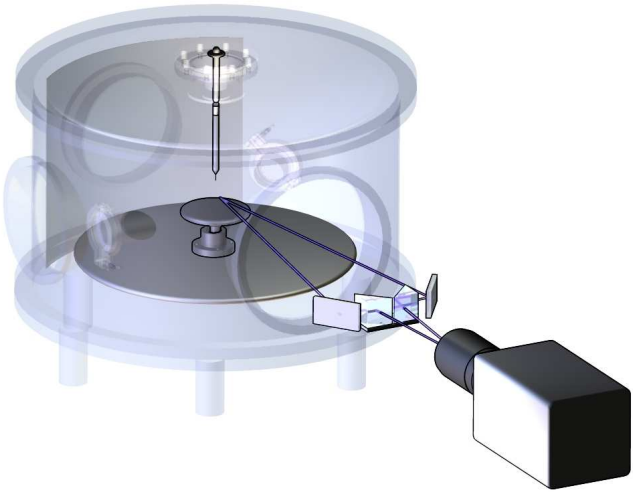}

\caption{\label{fig:SetupOverview}Schematic overview of the stereoscopic measurement
set-up with a schematic drawing of the two image paths. In this example
an anode with a single tip is installed inside the discharge vessel.}

\end{figure}

All experiments presented here have been performed in ambient air
with a cathode-anode gap of 40~mm. Although the experiments have
been conducted inside a vacuum-vessel, in many cases the front window
of this vessel was removed so that the air was identical to the ambient
air inside the laboratory. In cases where pressures below 1000~mbar
were used, the front window was mounted and the vessel contents were
flushed with ambient air. The same electrical circuit as described
in~\cite{Nijdam2008} was used. With this circuit, a positive (cathode
directed) corona discharge is created by discharging a capacitor with
a fast switch. We use a spark-gap as switch. This results in a voltage
pulse on a pointed tip or wire with a rise-time of about 20~ns, a
maximum voltage between 6 and 55~kV and a decay time of a few microseconds.
More information about this circuit and the discharge vessel can be
found in~\cite{Briels2006} and~\cite{Briels2008b}.

\subsection{(Stereo) imaging and image evaluation}

The corona discharge produced by the circuit is imaged onto an intensified
CCD-camera (Stanford Computer Optics 4QuickE) with a minimum gate
time of 2~ns. The original camera frames of all images are 1360 by
1024 pixel, 14~bit gray-scale images. All camera images presented
in this paper are false-colour representations of these original camera
frames. Image processing and measurements (e.g. of streamer diameters)
were done on the original full resolution 14~bit gray-scale frames.

In most experiments, an optical system was positioned between the
camera and the discharge in order to enable stereo photography. This
optical system consists of two prisms and two mirrors as shown in
figure~\ref{fig:SetupOverview}. With this set-up we are able to
produce stereoscopic images, as can be seen in figures~\ref{fig:onetipStereoReconnection},
\ref{fig:twotipsasymStereoReconnection}, \ref{fig:twotipsasymStereoReconnectionDelayed}
and~\ref{fig:Wire-plateNoMerging}. In order to reconstruct the 3D
structure of (a part of) a streamer discharge, custom built software
is used in which a line segment is placed manually over identical
streamers in both left and right views. The xyz coordinates of the
end points of this line are now calculated with a method described
below.

Because we used a 40~mm gap instead of the 140~mm gap discussed
in~\cite{Nijdam2008}, the camera, prisms and mirrors have been placed
closer to the vacuum vessel. The full angle between the two optical
paths is 10$\textdegree$ in all stereo measurements presented here.

\begin{figure}
\includegraphics[width=16cm]{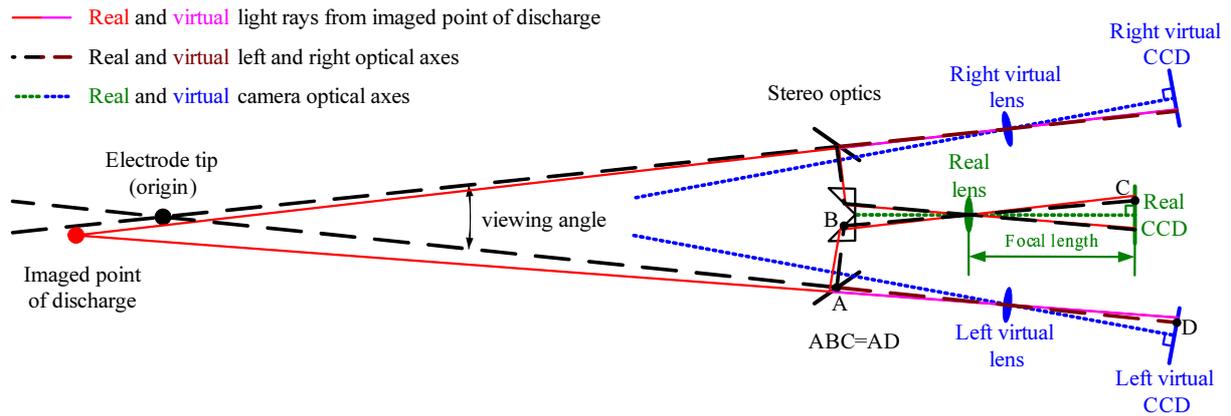}

\caption{\label{fig:ReconstructionOptics}Schematic overview (top view) of
the stereoscopic reconstruction technique. Both the real optics and
camera and the two virtual cameras are shown. The imaged point is
imaged on the two virtual CCD's and the real CCD. It can be seen that
the distance between the optical axis and the imaged point on the
CCD differs on the left and the right hand side. In the actual reconstruction,
the third dimension is also used.}

\end{figure}

Another difference between the method described in~\cite{Nijdam2008}
and the current method is the improved geometric reconstruction method.
The present 3D reconstructions are obtained by using a complete 3D
geometrical computation assuming a pin-hole camera. In this method,
two virtual cameras are placed so that they will produce the left
and right images without the prisms and mirrors as shown in figure~\ref{fig:ReconstructionOptics}.
In other words, the distance between these cameras and the origin
(on the vertical axis of the vacuum vessel) is equal to the total
path length of one of the two paths drawn in figure~\ref{fig:SetupOverview}.
The angle between the two virtual cameras and the origin is equal
to the angle between the two paths (10$\textdegree$ in our case).

Both virtual cameras are represented as pinhole cameras; they consist
of an infinitely small hole and a screen (the ICCD). In order to reconstruct
a 3D location of a point, the vector from the pixel location of this
point to the pinhole location is calculated for both cameras. The
crossing point of these two vectors now determines its real 3D location.
In real life measurements the vectors will of course never really
cross in full 3D; therefore the point halfway between their closest
points is used.

The assumption of a pinhole camera neglects lens artifacts like chromatic
aberration and pincushion or barrel distortion. The earlier evaluations
in \cite{Nijdam2008} used a simpler model assuming two cameras far
away from the system with a very large focal length. The present method
reduces the absolute position errors of points far away from the origin
(here defined as the position of one of the tips or an arbitrary position
on the wire, see figure~\ref{fig:ReconstructionOptics}). The local
errors (relative position errors of two points close together) have
not changed significantly by employing the new method.

We estimate the error in absolute and relative position at around
2\% and 0.5\% of the vertical image size. In our set-up the vertical
image size is roughly 50~mm. This means that the absolute error is
about 1~mm and the relative error is about 0.3~mm.

If reconnection or merging occurs, it should be visible in both images
of a stereoscopic photo. Furthermore, the vertical position of the
merging/reconnection location should be the same in both images. If
this is clearly the case then it can be concluded that the merging
or reconnection really takes place. In many cases, this method has
been used instead of the more labor intensive complete 3D reconstruction.

In some cases we have measured the width of the streamer channels.
This is achieved with the following method: A line is drawn manually
along a straight section of a streamer channel in a 2D camera image.
Now a region is chosen that is large enough to easily encompass the
whole channel, but not so large that it covers neighbouring channels.
The pixel intensities of all pixels in this region are averaged along
the direction of the line. This leads to an averaged crosscut profile
of the streamer channel. From this profile, the base level is subtracted
and then the full width at half maximum (FWHM) is determined.

\subsection{Anode geometries}

\begin{figure}[h]
\includegraphics[width=12cm]{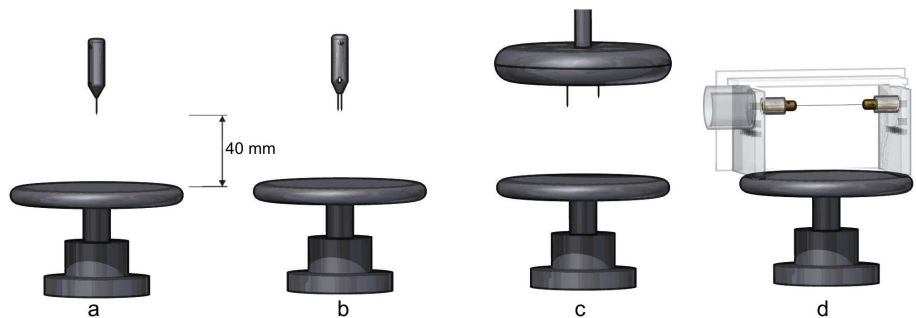}

\caption{\label{fig:AnodeGeometries}Schematic drawing of the four different
anode geometries above the cathode plane. a)~single tip; b)~symmetrical
double tips; c)~asymmetrical double protrusion from a plane; d)~wire.
In the image of the wire geometry, all non-metal components are shown
as semi-transparent. The drawing is to scale.}

\end{figure}

Four different anode geometries have been used; they are shown in
figure~\ref{fig:AnodeGeometries}: (a)~a single tip, (b)~symmetrical
double tips, (c)~asymmetrical double protrusions from a plane and
(d) a wire. The difference between the double tip geometry~(b) and
the double protrusion geometry~(c) is that in the double tip geometry,
both tips extend from a pointed holder, while in the double protrusion
geometry, the tips extend from a plane similar to the cathode plate.
In this geometry, the background electric field at some distance from
the protrusions is quite similar to a plane-plane geometry; it is
quite homogeneous in contrast to a point plane geometry. This protrusion
geometry~(c) is roughly similar to a needle-array electrode~\cite{Takaki2005,Krasnochub},
but it has only two needles. In particular, we will use an asymmetric
configuration where one tip extends significantly further out of the
plane than the other.

The distance between the tip(s) or the wire and the cathode plate
was always set to 40~mm. In the asymmetric geometry~(c), 40~mm
is the distance between the tip of the longer protrusion and the plane.
The same tips have been used in all geometries, made of 1~mm diameter
tungsten rods, with a conical pointed end. The tip radius is about
15~$\mu\mathrm{m}$. The wire in the wire geometry is a 0.2~mm diameter
kanthal wire.

\subsection{Timings and delays}

\begin{figure}
\includegraphics[width=8.5cm]{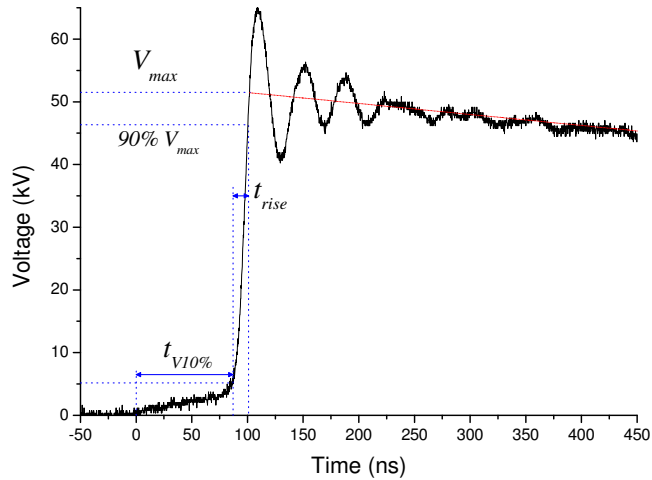} 

\caption{\label{fig:Voltage-curve}Typical voltage curve of streamer discharge
with definitions of important characteristics.}

\end{figure}

In order to properly correlate the timing of camera images to the
voltage curve, a full analysis of all delays in our experimental set-up
has been conducted. With the results of this analysis we can define
an absolute timescale for all measurements. We define the origin of
the time scale ($t=0$~ns) as the moment when the voltage imposed
on the anode starts to increase above 0~V. This is illustrated in
figure~\ref{fig:Voltage-curve}. \label{Section:Timings}

The maximum voltage ($V_{max}$) is defined as the crossing point
of the increasing voltage slope with a linear fit of the beginning
of the decreasing voltage slope. In the case of figure~\ref{fig:Voltage-curve}
this leads to $V_{max}=52\,\mathrm{kV}$. The duration of the slow
voltage increase just after $t=0$~ns (here about 80~ns) depends
mainly on the pressure in the spark gap switch and, to a lesser extent,
on the streamer vessel pressure and anode geometry. Therefore we have
defined another important point: $t_{V10\%}$, the time when the voltage
has reached $V=0.1\cdot V_{max}$. The rise-time ($t_{rise}$) of
the voltage slope is also important. We have defined it as the time
between $V=0.1\cdot V_{max}$ and $V=0.9\cdot V_{max}$. These characteristic
values of the voltage pulse are influenced by resistors in the power
supply and the impedance of the discharge and therefore vary between
experiments.

The camera delay ($t_{start}$) is related to the same origin of the
time scale as the voltage pulse. The camera gate is opened at $t_{start}$
and stays open for a specified time ($t_{gate}$). A more detailed
discussion of the timing scheme can be found in the appendix.

\section{Experimental results}

\subsection{Reconnection}

\begin{figure}
\includegraphics[width=8cm]{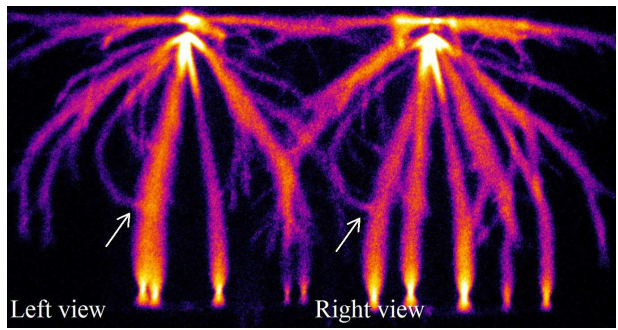}

\caption{\label{fig:onetipStereoReconnection}A stereo image of a streamer
reconnection in single tip anode geometry~(a). The two views of the
single tip event overlap a bit in the middle of the figure. The most
striking reconnection location is marked with an arrow in both views.
Experimental settings: gas fill: 1000~mbar ambient air; $V_{max}=52\,\mathrm{kV}$;
$t_{V10\%}=87\,\mathrm{ns}$; $t_{rise}=24\,\mathrm{ns}$; $t_{start}=52\,\mathrm{ns}$;
$t_{gate}=50\,\mathrm{ns}$. The voltage curve in this experiment
is very similar to the one shown in figure~\ref{fig:Voltage-curve}.
As can be seen from the timing parameters of this experiment, the
complete image is shot before the voltage pulse reached its maximum.}

\end{figure}

\begin{figure}
\includegraphics[width=9cm]{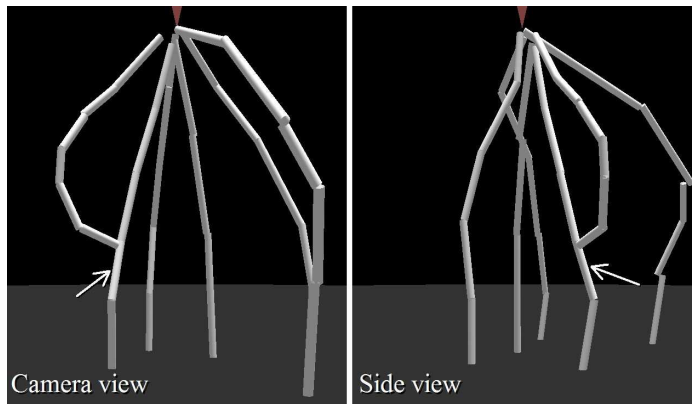}

\caption{\label{fig:onetip3DReconnection}Orthogonal views of the 3D~reconstruction
of the reconnection event from figure~\ref{fig:onetipStereoReconnection}.
Again the reconnection location is marked with an arrow in both views.
Not all streamer channels from the original images are represented
in this reconstruction. An animation of this reconstruction can be
found online as figure6.avi ($\sim$2MB) with the marked reconnection
event shown in blue.}

\end{figure}

Streamer reconnection is studied in the single tip geometry~(a) and
in the double protrusion geometry~(c) (see figure~\ref{fig:AnodeGeometries}).
Apparent reconnections in the single tip geometry~(a) were observed
and discussed before by Briels et al.~\cite{Briels2006}, based on
normal photography. A stereo image of a single tip streamer discharge
with apparent reconnection(s) is shown in figure~\ref{fig:onetipStereoReconnection}.
A partial 3D reconstruction of this discharge is shown in figure~\ref{fig:onetip3DReconnection}.

The stereo image already shows that reconnection occurs in both the
left and right hand views at the same vertical position. This indicates
that there is indeed a thin streamer channel reconnecting to a thick
streamer channel. This is confirmed in the 3D reconstruction. It was
found that reconnections as the one shown are observed in about 50\%
of the images taken under the conditions of figure~\ref{fig:onetipStereoReconnection}.
Because some images are not very clear and in some cases reconnecting
streamer channels are obscured by other streamer channels, we conclude
that reconnections occur in the majority of discharge events under
these conditions.

\begin{figure}
\includegraphics[width=8cm]{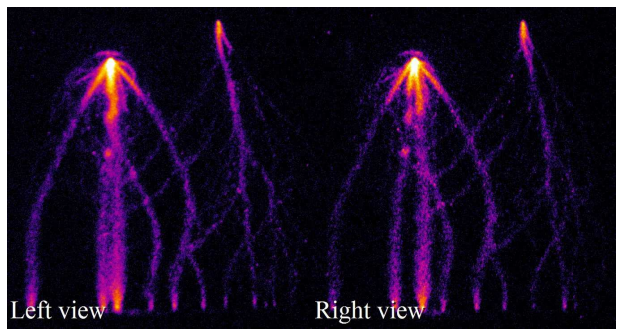}

\caption{\label{fig:twotipsasymStereoReconnection}A stereo image of multiple
streamer reconnections in the double protrusion-plane anode geometry
(c). The horizontal distance between the two tips is 18~mm, the left
tip protrudes 14~mm from the plane and the right tip protrudes 8~mm
from the plane. Other experimental settings: gas fill: 1000~mbar
ambient air; $V_{max}=50\,\mathrm{kV}$; $t_{V10\%}=15\,\mathrm{ns}$;
$t_{rise}=35\,\mathrm{ns}$; $t_{start}=10\,\mathrm{ns}$; $t_{gate}=100\,\mathrm{ns}$.}

\end{figure}

\begin{figure}
\includegraphics[width=10cm]{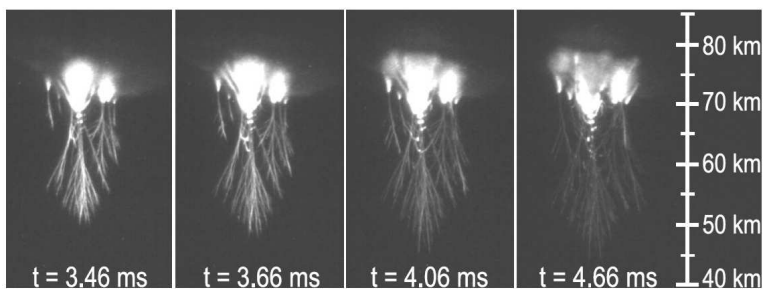}

\caption{\label{fig:Sprite_Cummer}High speed sprite discharge images from
13 August 2005 at 03:12:32.0 UT showing apparent reconnection between
different channels. These discharges occur above thunderclouds. The
picture is reproduced from \cite{Cummer2006}, copyright 2006 American
Geophysical Union. Reproduced/modified by permission of American Geophysical
Union.}

\end{figure}

Streamer reconnections can also be observed in the double protrusion
anode geometry~(c). In this geometry, the thickest and earliest streamer
channels originate from the tip that protrudes farthest from the plane.
An example of such a discharge event is shown in figure~\ref{fig:twotipsasymStereoReconnection}.
This image shows multiple reconnections from streamer channels originating
from the right tip to streamer channels originating from the left
tip. (We continue to use the term reconnection as these streamers
originate from the same electrode.) All these reconnections are clearly
visible in both views and are therefore interpreted as real reconnections.
The width (full width at half maximum, FWHM) of the thick, early channels
is about 1.1~mm, the width of the thin channels about 0.6~mm. In
figure~5 of~\cite{Briels2008}, Briels \textit{et al.} have reported
similar values for the width of streamers created under these conditions.

Reconnections as shown in figures~\ref{fig:onetipStereoReconnection}
and~\ref{fig:twotipsasymStereoReconnection} are observed frequently
under these experimental conditions. We find that reconnection only
occurs to streamer channels that have crossed the entire gap and end
on the cathode plate.

In figure~\ref{fig:Sprite_Cummer}, a sprite discharge is shown that
remarkably resembles the streamer discharge shown in figure~\ref{fig:twotipsasymStereoReconnection}.
Similar reconnection events are visible although it can not be proved
that they are real because no stereo-photographic images are available.
The similarities and differences between sprite and streamer reconnections
will be discussed in more detail in section~\ref{sec:ConclusionsAndDiscussion}.

\begin{figure}
\includegraphics[width=8cm]{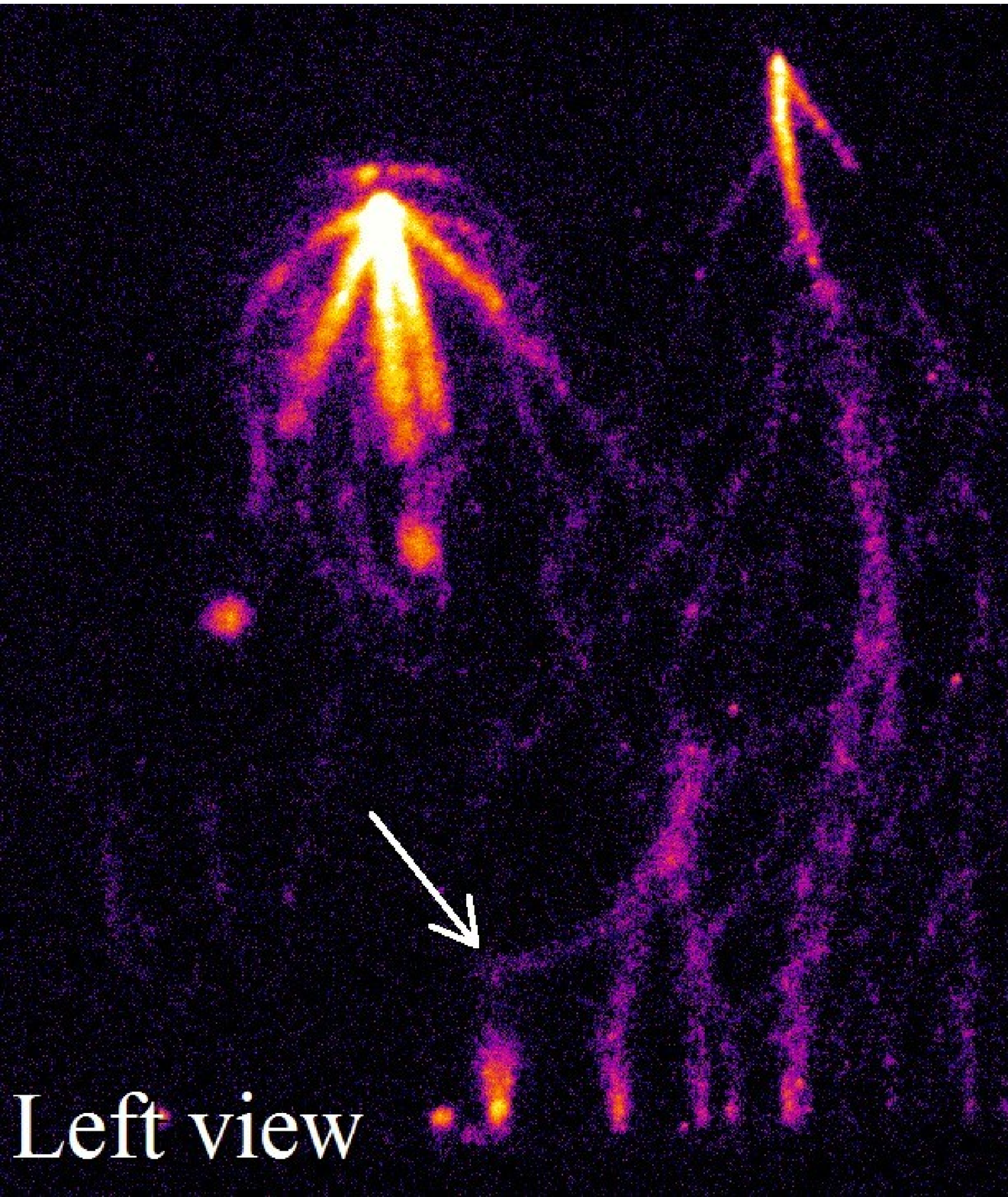} \includegraphics[height=4.79cm]{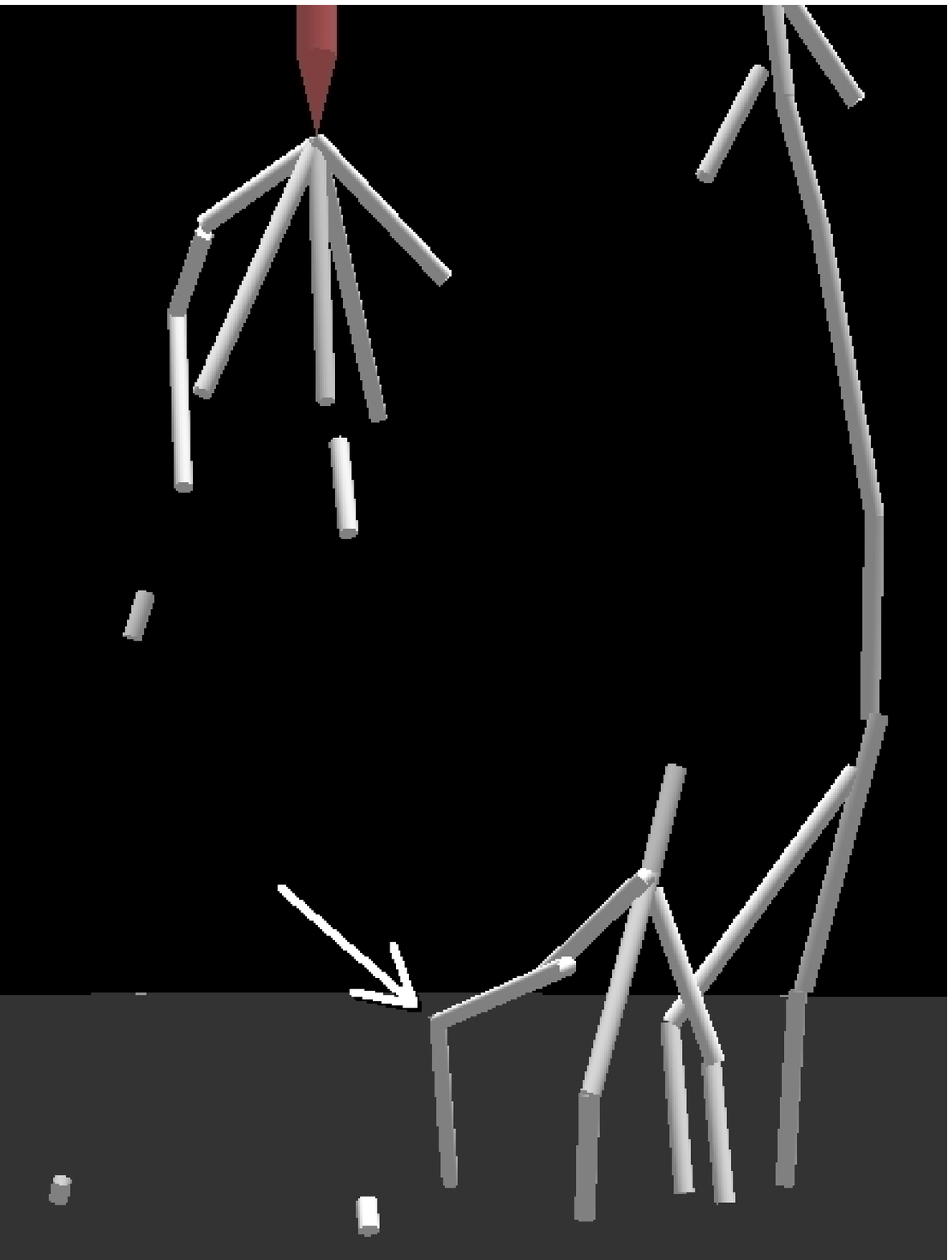}

\caption{\label{fig:twotipsasymStereoReconnectionDelayed}A stereo image and
its 3D reconstruction of multiple streamer reconnections in the double
protrusion-plane anode geometry (c). The experimental conditions are
the same as in figure~\ref{fig:twotipsasymStereoReconnection} except
for the longer camera delay $t_{start}=50\,\mathrm{ns}$. Note that
the streamer channel indicated with the arrow makes a nearly 90$\textdegree$
turn. An animation of the reconstruction can be found online as figure9.avi
($\sim$2MB) with the indicated turn shown in blue.}

\end{figure}

Briels \emph{et al. }\cite{Briels2008b} have reported that thicker
streamer channels are always faster than thin streamer channels. In
our experiments, tens to hundreds of nanoseconds after the thick channels
have bridged the gap, thinner and slower streamers can connect to
the conducting traces left behind by the early thick streamers. This
has been investigated by increasing the delay of the camera so that
only the late streamers are visible. An example of such an image can
be seen in figure~\ref{fig:twotipsasymStereoReconnectionDelayed}.
It shows that most of the length of the thick channels seen in figure~\ref{fig:twotipsasymStereoReconnection}
is no longer visible. Only the upper part (secondary streamers) and
the cathode spots of these channels remain clearly visible. This can
easily be understood by the fact that only the propagating head of
a streamer channel emits a significant amount of radiation; therefore
the channel behind the head is usually invisible or very dim after
the head has passed \emph{cf.} figure~1 in~\cite{Ebert2006a}.

A striking feature in figure~\ref{fig:twotipsasymStereoReconnectionDelayed}
is the streamer channel indicated with the arrow. This streamer channel
seems to change direction instantaneously by about 90$\textdegree$.
These direction changes are observed in most images with longer delays
under these experimental conditions. They appear very similar to the
shape of the streamers reconnecting as in figure~\ref{fig:twotipsasymStereoReconnection}.
Therefore, these direction changes are interpreted as streamer reconnection.
This confirms that reconnection is indeed the attraction of a late
streamer channel towards an earlier streamer channel.

\subsection{Merging}

\begin{figure}
\includegraphics[width=8cm]{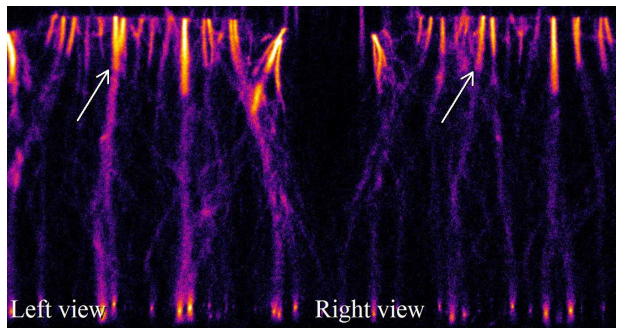}

\caption{\label{fig:Wire-plateNoMerging}Stereo image of a wire-plate discharge.
A possible merging location in the left hand view is indicated with
an arrow. However, the right hand view clearly shows that in reality
no merging occurs. Experimental settings: gas fill: 1000~mbar ambient
air; $V_{max}=45\,\mathrm{kV}$; $t_{V10\%}=15\,\mathrm{ns}$; $t_{rise}=22\,\mathrm{ns}$;
$t_{start}=0\,\mathrm{ns}$; $t_{gate}=1000\,\mathrm{ns}$.}

\end{figure}

Streamer merging was first suggested as an interpretation of experiments
in a wire-plate discharge. 2D pictures of Grabowski \emph{et al.}~\cite{Grabowski2005}
and Winands \emph{et al.}~\cite{Winands2006} show possible merging
of streamer channels close to the wire. We have reproduced such experiments
with the wire-plate electrode geometry as shown in figure~\ref{fig:AnodeGeometries}d.
During this investigation, we have never found a definite case of
merging of streamer channels in the hundreds of discharge events studied.
Often channels seem to merge in one of the images, but are clearly
not merging in the other image. An example of such an event is shown
in figure~\ref{fig:Wire-plateNoMerging}. Here, in the left image
two streamer channels seem to merge. However, the right image clearly
shows that this is not the case.

The left image also shows that two channels propagate downwards from
the apparent merge location towards the cathode plate. This is already
an indication that no real merging occurs. In rare cases (less than
1\% of the images), the image quality around an apparent merging location
was not good enough to definitely conclude that no merging occurred,
but the propagation of multiple channels from these locations again
indicates that no merging occurs. The same conclusion is drawn from
measurements at 400~mbar ambient air.

The average separation between streamers initiating from the wire
in figure~\ref{fig:Wire-plateNoMerging} is close to 2~mm. This
value is in agreement with streamer observations in a wire-plane geometry
of Winands \emph{et al.} \cite{Winands2008} and Creyghton \emph{et
al.} \cite{Creyghton1994a}.

The streamer channels are more parallel in the wire-plate discharge
than in the (double) tip-plate discharges. This is because the electric
field lines diverge around the needle electrode in the projection
plane of the picture while they are parallel for the wire electrode.
However, in the direction perpendicular to the image plane, the streamers
do diverge significantly and show similar spatial distributions as
in the point-plane discharges. This divergence has been observed before
by Winands et al. (\cite{Winands2007}, figure~D6).

\begin{figure*}
\includegraphics[width=1\columnwidth]{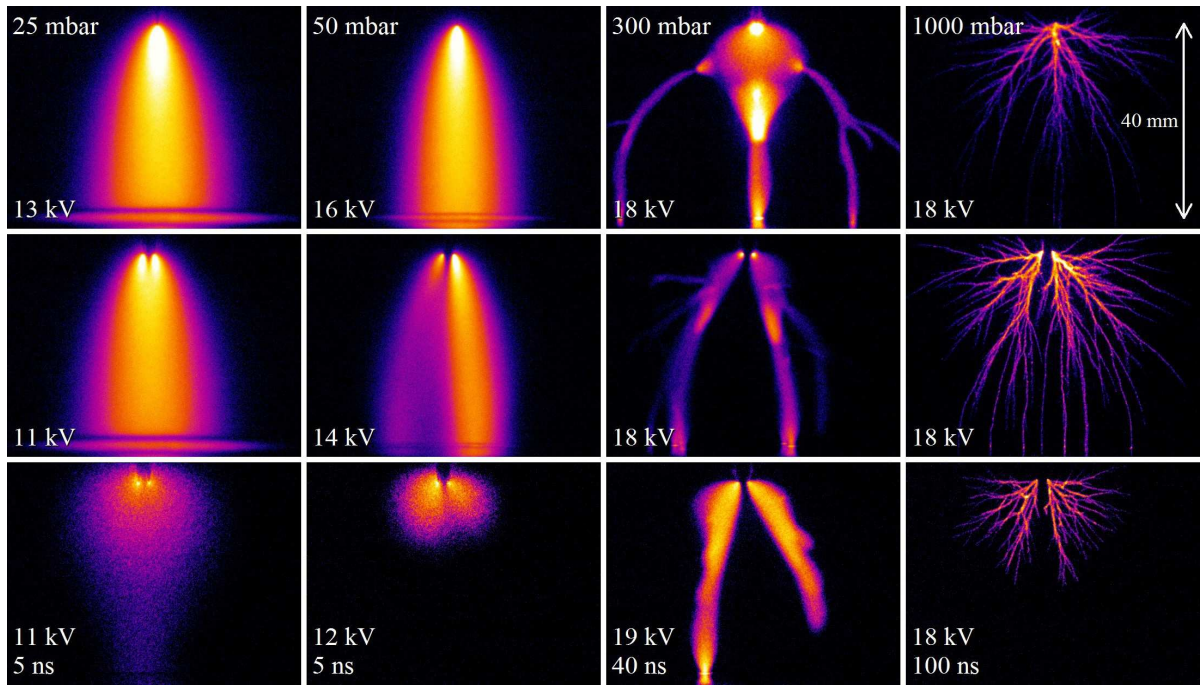}

\caption{\label{fig:merging8pics}Images of streamer discharges with the single
tip anode geometry~(a) in the top row, and with the symmetrical double
tip anode geometry~(b) in the middle and bottom rows. Within the
columns, the pressure varies from 25~mbar to 1000~mbar as indicated.
The two upper rows show time integrated images ($t_{gate}>1000\,\mathrm{ns}$),
the bottom row shows images with shorter gate times as indicated in
the panels. In all images $V_{max}$ is indicated. The horizontal
distance between the two tips is 2~mm. Voltage rise times varied
between 15 and 25~ns, the camera opens at $t_{start}\cong t_{V10\%}$.
The camera gain varies significantly between the images; discharges
are much brighter at lower pressures than at higher pressures.}

\end{figure*}

A more fundamental method to investigate streamer merging is to use
two tips close to each other as shown in figure~\ref{fig:AnodeGeometries}b.
In this case streamer channels originate from both of these tips simultaneously
(within 2~ns). These two streamer channels will now propagate in
a more or less parallel direction towards the cathode plate and may
merge or repel each other. Depending on pressure and other parameters,
the two original streamer channels may also branch once or more. The
chosen tip separation distance of 2~mm is similar to the average
distance between streamers in the wire experiments at 1~bar. In our
experiments both tips extend about 10~mm from the tip holder.

Images of streamer discharges originating from double tips are compared
to images of streamers originating from a single tip in figure~\ref{fig:merging8pics};
the pressure varies from 25~mbar to 1000~mbar between the columns;
the rows show single and double tip experiments with long exposure
times and double tip experiments with short exposure times. The figure
shows that for the double tip discharges, always two initial streamer
channels are formed. These two channels repel each other for all pressures,
except for 25~mbar where they merge. The width of the streamer (FWHM)
is 23~mm at 25~mbar. This is 11.5 times as large as the distance
between the two tips. At 50~mbar, the width of the single tip streamer
is still 18~mm, 9 times the tip distance, but the streamers do not
merge.

In cases when one of the streamers initiates shortly before the other
(by any reason, randomly or systematically), it can shield the second
streamer from the background electric field. This causes the second
streamer to be less powerful than the first streamer. This can be
observed in the 50~mbar double tip images. Here, it is clear that
one streamer channel dominates the discharge and the other channel
moves away from the central vertical axis. Which tip or side is dominant
changes randomly from discharge to discharge and in some cases both
channels are equally bright and symmetrically shaped. That one side
often dominates the discharge, is observed for pressures between 50
and 250~mbar.

The images of figure~\ref{fig:merging8pics} are taken with peak
voltages of around 12~kV for the 25~mbar images to 18~kV for the
300~mbar and 1000~mbar images%
\footnote{It was attempted to use a fixed $V_{max}$ of 18~kV for all images.
However, for pressures below 100~mbar, $V_{max}$ had to be decreased
to prevent sparking in the top feedthrough of the set-up.%
}. Also for other voltages the same merging behaviour as function of
pressure was observed. At 25~mbar, merging still occurs with 6~kV
peak voltage, while at higher pressures no merging is observed at
both lower and higher voltages than the ones shown in figure~\ref{fig:merging8pics}.

\section{Discussion and conclusions}

\label{sec:ConclusionsAndDiscussion}From the images shown above and
from all other images taken during these investigations, it can be
concluded that reconnection of late streamer channels on earlier streamer
channels occurs frequently in streamer discharges with many streamer
channels. It is suggested~\cite{Briels2006} that the reconnection
is caused by electrostatic attraction of a late streamer to a conducting
channel left by an early streamer that already has reached the oppositely
charged electrode and has changed polarity. The late streamer approaches
the early channel almost perpendicularly --- like an electrode plate
or wire --- which is another argument in favor of an electrostatic
mechanism. In order to study reconnection of streamers theoretically,
a complete three dimensional model would be required. Such streamer
simulations, however, are only in the first stages of development~\cite{Luque2008b,Luque2008a}.

Besides electrostatic attraction, two other interaction mechanisms
between streamers can be imagined as the reason for reconnection:
magnetic attraction and photo-ionization. However, magnetic attraction
between current channels only occurs when these channels are more
or less parallel and would not lead to the near perpendicular reconnections
that are observed. Besides, the currents in these streamer channels
are low and would not lead to any significant Lorentz force. Photo-ionization
can also be excluded because decays exponentially like $e^{-\nicefrac{r}{\ell}}$
for distances $r$ larger than the photo-ionization length $\ell=1.6\,\mathrm{mm}$
(at atmospheric pressure according to \cite{Penney1970,Zhelezniak1982}),
while electrostatic attraction decays as $\frac{1}{r^{2}}$. Therefore
photo-ionization is much weaker at distances exceeding $\ell$, and
can not turn the streamer path over large distances as in figures~\ref{fig:onetipStereoReconnection}-\ref{fig:twotipsasymStereoReconnection}
and~\ref{fig:twotipsasymStereoReconnectionDelayed}.\\

The reconnections in sprites~\cite{Cummer2006} show a very similar
signature, as a comparison of Figs.~\ref{fig:Sprite_Cummer} and
\ref{fig:twotipsasymStereoReconnection} shows: the head of a late
sprite streamer is attracted to an earlier formed channel. However,
no stereoscopic imaging is available in this case to decide whether
the effect is real.

The similarity of sprites and streamers by now is well established,
see~\cite{Briels2008b,Pasko2007,Stenbaek-Nielsen2008}. However,
there is another important difference, namely the early attracting
sprite channel is not connected to some electrode to explain its polarity
change. A possible mechanism is here a charge separation along the
sprite streamer channel; this hypothesis should be investigated in
future theoretical work.\\

Merging of more or less parallel, simultaneously propagating streamer
channels was only observed at low pressures (25~mbar), but not at
higher pressures under the conditions of these experiments (overvolted
40~mm gaps, ambient air, 2~mm or more tip separation). Only when
the streamer diameter (FWHM) is at least 10 times larger than the
tip separation, we observe merging. All observations show that thinner
streamer heads always do repel each other and will remain separate
while propagating between anode and cathode.

A mechanism for streamer merging was recently proposed by Luque \textit{et
al.}~\cite{Luque2008b}; as photo-ionization in air is a nonlocal
ionization reaction acting over a reduced length scale of about 1.6~mm~bar
at normal temperature, the ionization cloud around the streamer heads
decays smoothly over this length scale. (The unit mm~bar is used
because the length scales of streamers are expected and observed to
scale quite well with pressure \cite{Ebert2006a,Briels2008b}.) When
the streamer heads get so close that their photoionization zones overlap,
impact ionization can further enhance the ionization in the area between
the heads and the streamer heads can merge. Again,\textit{ }magnetic
attraction of parallel current carrying channels is very weak because
of the low currents and can therefore be excluded.

We can compare the experimental results with the simulations of Luque
\textit{et al.}~\cite{Luque2008a} on streamer merging due to photo-ionization.
In these calculations, the pressure $p$ was varied while the reduced
seed distance was fixed to $p\cdot d=230\,\mu\mathrm{m\cdot bar}$
at room temperature, here $d$ is the seed separation. Up to $p=1$~bar,
the streamers in air always merged in these simulations. In the double
tip measurements presented here, the real tip distance was fixed,
therefore the reduced tip distance ranges from $50\,\mu\mathrm{m\cdot bar}$
to $2\,\mathrm{mm\cdot bar}$. At 115~mbar, the 2~mm tip separation
gives a reduced tip separation of $230\,\mu\mathrm{m\cdot bar}$,
but the streamers repel each other; they merge only below $p\cdot d=100\,\mu\mathrm{m\cdot bar}$.
However, the streamers in the theoretical work emerge from an avalanche
in a homogeneous electric field, while the streamers in the present
experiments emerge from two needle electrodes. Therefore a discrepancy
by a factor of 2 or more is not unreasonable, and we conclude that
experiments and theory need to be developed further before they can
be compared quantitatively. Furthermore, the theoretical prediction
relies on the photo-ionization lengths that were measured by Penney
and Hummert in 1970~\cite{Penney1970} and whose accuracy is widely
doubted \cite{Pancheshnyi2005,Luque2007,Nudnova2008}; but no other
data are available.

Streamers in experiments are never initiated exactly simultaneously.
We observe that inception of the different streamer channels occurs
within 2~ns but we can not be certain that there is no jitter on
smaller timescales. In the double tip experiments we can be certain
that there is not one favorite tip because the dominant streamer changes
from left to right and back randomly between discharge events. We
can however, not be certain that there is no stochastic jitter in
the inception times. This can lead to two streamers with a small mutual
delay that also could prevent merging. If this is the case, it will
probably be always present in real world experiments. Numerical simulations
can answer if a time delay of e.g. 1~ns between two streamers will
prevent merging. Such a time delay between inception can also be responsible
for the appearance of a dominant and a weaker streamer channel; the
streamer that is initiated first shields the second streamer from
the background electric field.

In wire-plate discharges, apparent streamer merging is occasionally
observed on 2D-images. However, in these discharges we have never
observed merging with any degree of certainty with stereo-photography.

\ack{}{S.N. acknowledges support by STW-project 06501, part of the Netherlands'
Organization for Scientific Research NWO.}

\appendix

\section*{Appendix. Timing scheme of the experiment}

\setcounter{section}{1}

\begin{figure}
\includegraphics[width=16cm]{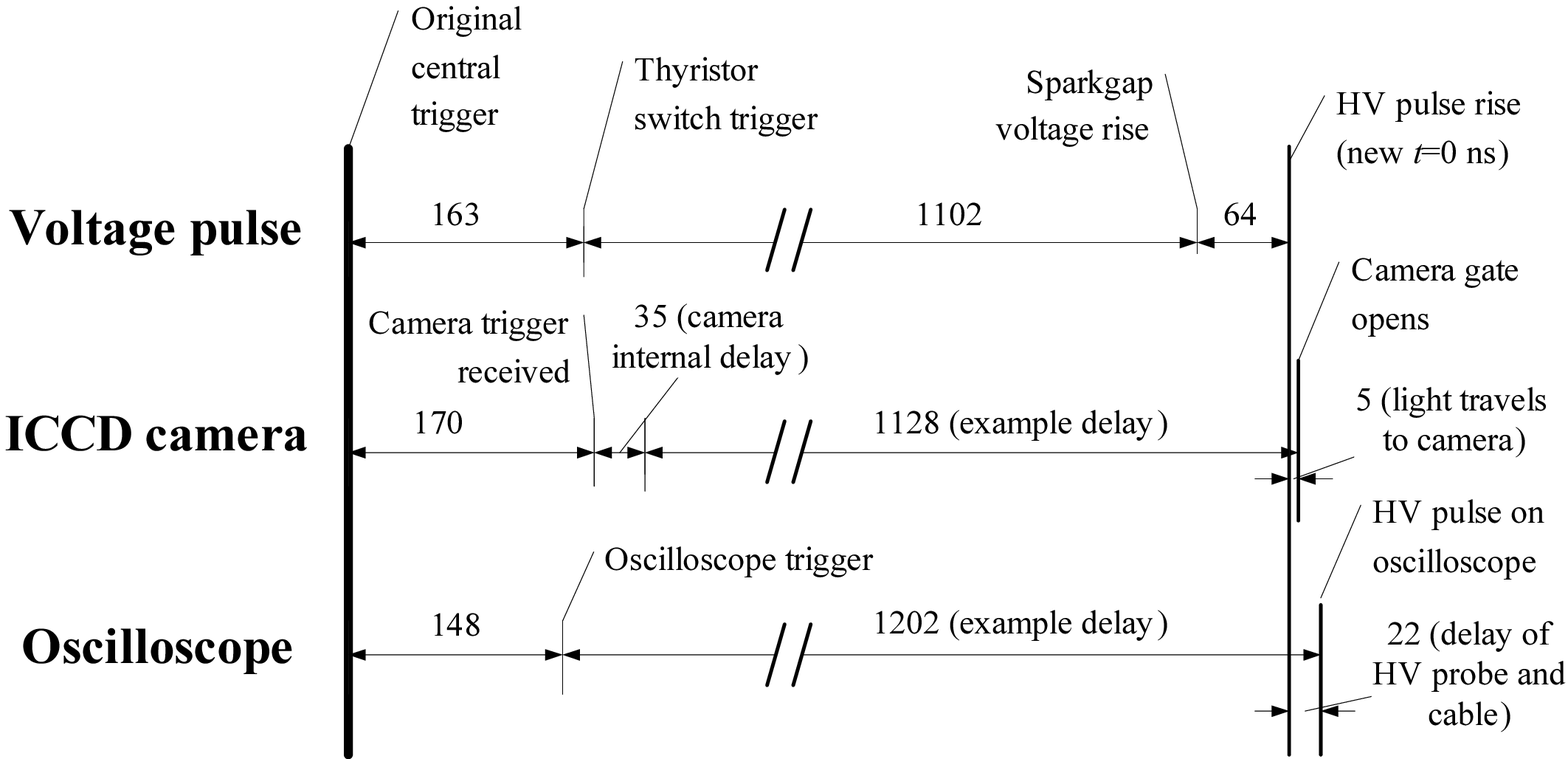}

\caption{\label{fig:timingscheme}Simplified timing scheme. All values are
times in ns.}

\end{figure}

In order to determine the timing characteristics of the set-up, a
full analysis of all timings and delays has been performed. In this
analysis, the delays of all cables, optical fibres, probes and other
equipment has been determined and combined. A sketch is shown in figure~\ref{fig:timingscheme}.
In this figure, three time paths for different parts of the set-up
are shown. These three time paths are all initiated by a central trigger
from a function generator. Details of the time paths will be discussed
below.

\subsubsection*{Voltage pulse}

From the central trigger, the trigger pulse is fed through coaxial
cables and an optical fibre to the thyristor switch in the high voltage
circuit. This takes 163~ns. The thyristor switch in turn triggers
the spark-gap 1102~ns later. The spark-gap then causes the rise of
the high voltage pulse on the anode with a 64~ns delay. All three
sections in this time-path have fixed delays with jitter of maximum
2-3~ns. These delays are intrinsic properties of the cables and equipment.

\subsubsection*{ICCD Camera}

The ICCD camera is also triggered from the central trigger through
coaxial cables and an optical fibre. This takes 170~ns. After receiving
this trigger, the camera has an internal delay of 35~ns before it
can open the gate. From this moment, the gate is opened after a user-specified
delay. In the example from figure~\ref{fig:timingscheme}, this delay
is set to 1128~ns in order to capture the beginning of the voltage
pulse and the streamer initiation. Because of the short time-scales
involved, we must account for the path-length of the light rays traveling
from the discharge to the camera.

\subsubsection*{Oscilloscope}

Again, the oscilloscope is triggered from the central trigger through
coaxial cables and an optical fibre. This takes 148~ns. The rise
of the high voltage pulse will occur some time after this trigger;
in our example 1180~ns. Because the oscilloscope is connected to
a high voltage (HV) probe with a long coaxial cable, the voltage rise
will be detected on the oscilloscope 22~ns later.

\subsection*{Resulting timing}

By using the known and set delays of the set-up, we can redefine a
new origin of the time-scale that is the same for both the oscilloscope
and the camera as is explained in section~\ref{Section:Timings}.
In the example from figure~\ref{fig:timingscheme}, we have to subtract
74~ns from the time as shown on the oscilloscope to convert this
to the time-scale of the camera. In other words, an event occurring
at the new $t=0\,\mathrm{ns}$ will be registered on the oscilloscope
at 1128~ns and on the camera at 1202~ns.

\section*{References}

\bibliographystyle{unsrt}
\bibliography{reconnections}

\end{document}